\documentclass[8pt]{article}
\usepackage[a4paper,margin=1in]{geometry}
\usepackage{latexsym}
\usepackage{graphicx}
\usepackage{wrapfig}
\usepackage{caption} 
\usepackage{lipsum}
\usepackage{float}
\usepackage{mathtools}
\usepackage{cuted}
\usepackage{amsmath,mathtools}
\usepackage{graphicx}
\usepackage{nccmath}
\usepackage{amsfonts,amssymb,amsmath}
\usepackage{hyperref}
\usepackage{xcolor}
\usepackage[ruled,vlined]{algorithm2e}
\usepackage[normalem]{ulem}
\usepackage{cite}
\usepackage{setspace}
\providecommand{\keywords}[1]
{
  \small	
  \textbf{\textit{Keywords---}} #1
}

\usepackage{scalerel}
\usepackage{tikz}
\usetikzlibrary{svg.path}

\definecolor{orcidlogocol}{HTML}{A6CE39}
\tikzset{
  orcidlogo/.pic={
    \fill[orcidlogocol] svg{M256,128c0,70.7-57.3,128-128,128C57.3,256,0,198.7,0,128C0,57.3,57.3,0,128,0C198.7,0,256,57.3,256,128z};
    \fill[white] svg{M86.3,186.2H70.9V79.1h15.4v48.4V186.2z}
                 svg{M108.9,79.1h41.6c39.6,0,57,28.3,57,53.6c0,27.5-21.5,53.6-56.8,53.6h-41.8V79.1z M124.3,172.4h24.5c34.9,0,42.9-26.5,42.9-39.7c0-21.5-13.7-39.7-43.7-39.7h-23.7V172.4z}
                 svg{M88.7,56.8c0,5.5-4.5,10.1-10.1,10.1c-5.6,0-10.1-4.6-10.1-10.1c0-5.6,4.5-10.1,10.1-10.1C84.2,46.7,88.7,51.3,88.7,56.8z};
  }
}

\newcommand\orcidicon[1]{\href{https://orcid.org/#1}{\mbox{\scalerel*{
\begin{tikzpicture}[yscale=-1,transform shape]
\pic{orcidlogo};
\end{tikzpicture}
}{|}}}}

\usepackage{hyperref}

\begin{document}

\title{Latency-Aware Offloading in Integrated Satellite Terrestrial  Networks}

\author{Wiem Abderrahim \orcidicon{0000-0001-5896-2307}, Osama Amin \orcidicon{0000-0002-0026-5960} ,  Mohamed-Slim Alouini \orcidicon{0000-0003-4827-1793} and  Basem Shihada\orcidicon{0000-0003-4434-4334} \\
\thanks{This work was supported in part by the Center of Excellence for NEOM Research at KAUST. The authors are with the  Computer, Electrical and Mathematical Sciences and Engineering (CEMSE) Divison, King Abdullah University of Science and Technology (KAUST), Thuwal 23955, Makkah Prov., Saudi Arabia (e-mail: {wiem.abderrahim, osama.amin, slim.alouini, basem.shihada}@kaust.edu.sa)}}
\date{}
\maketitle

{\begin{abstract}  Next-generation communication networks are expected to integrate newly-used technologies in a smart way to ensure continuous connectivity in rural areas and to alleviate the traffic load in dense regions. The prospective access network in 6G should hinge on satellite systems to take advantage of their wide coverage and high capacity. 
However, adopting satellites in 6G could be hindered because of the  {additional latency introduced}, which is not tolerable by all traffic types. Therefore, we propose a traffic offloading scheme that integrates both the satellite and terrestrial networks  to smartly allocate the traffic between them while satisfying different traffic requirements. Specifically, the proposed scheme offloads the Ultra-Reliable Low Latency Communication (URLLC) traffic to the terrestrial backhaul to satisfy its stringent latency requirement. However, it offloads the enhanced Mobile Broadband (eMBB)  {traffic to the satellite} since eMBB needs high data rates but is not always sensitive to delay. Our scheme is shown to reduce the transmission delay of URLLC packets, decrease the number of dropped eMBB packets, and hence improve the network's availability. Our findings highlight that the inter-working between satellite and terrestrial networks is crucial to mitigate the expected high load on the limited terrestrial capacity. \\\keywords{
  Enhanced Mobile Broadband (eMBB), Integrated Satellite Terrestrial  Networks (ISTN), Traffic offloading, Ultra-reliable low latency communication (URLLC).
}\end{abstract}

\section{Introduction}
With the approaching advent of the 5th generation of mobile cellular networks (5G), the next generation of mobile cellular networks (6G) is already required to afford more sophisticated features in terms of coverage and capacity \cite{dang2020what}. 6G should not only ensure global connectivity by connecting under-served areas where access to the Internet is limited or absent, but it should also mitigate the limitations in the capacity of terrestrial links, especially with the increasing data demand in dense networks. Of the mix of access technologies in 6G, satellite communications will play  {a pivotal role due to their wide coverage.}  {Notably, the new generations of low earth orbit (LEO) satellites are gaining a high-visibility in both industry and academia  because  they guarantee higher capacities (almost 1 Tbps \cite{del2019technical,de2017network}) and lower delays (20-50 milliseconds \cite{del2019technical}). In industry, different mega-constellations of LEOs  are being manufactured and partially launched into space lately. For example, the aerospace manufacturer SpaceX has an ambitious project to cover the Earth with around 12000 LEO satellites \cite{del2019technical}. SpaceX is racing against Amazon and OneWeb, which plan to launch 3236  satellites (Project Kuiper) and 650 satellites, respectively \cite{online1,del2019technical}. }They all aim to widen the terrestrial network’s connectivity and, more crucially to extend its limited capacity \cite{abo2019survey}.

 {In academia,  researchers also are exploring different approaches to extend the limited capacities of terrestrial networks \cite{liu2018space}. Recently, several efforts focused on traffic offloading in integrated satellite and terrestrial networks (ISTN)   \cite{liu2018space, zhang2019satellite, wang2018computation, cheng2019space, boya, li2020energy}.} For instance, the authors of \cite{zhang2019satellite} addressed the offloading problem in satellite mobile edge computing (SMEC). They studied and compared  different offloading locations, such as the terrestrial station, the LEO satellite, and the terrestrial gateway in SMEC  \cite{zhang2019satellite}.  The terrestrial gateway was the recommended offloading alternative because it guarantees lower complexity in terms of coverage and maintenance. In \cite{wang2018computation}, a double edge computing architecture was considered where the edge servers were deployed in satellite and terrestrial networks. Moreover, a cost matching algorithm was proposed to allocate the suitable edge servers to the offloaded tasks while optimizing the energy consumption and the offloading delay. In \cite{cheng2019space}, a  space-air-ground integrated network (SAGIN) architecture was studied where the aerial network nodes served as flying edge servers in the air segment and the LEO satellites in the space segment connected the Internet of things (IoT) devices with the cloud servers. SAGIN architecture helps the remote IoT applications to decide on the typical offloading location, namely locally, at the air segment, or at the space segment.    {In \cite{boya}, a traffic offloading scheme in ISTN was proposed to maximize the number of the accommodated users and their sum rate under a dynamic backhaul capacity constraint. The proposed scheme schedules the resources available in both networks to optimize users' association and to increase the dynamic backhaul capacity.}   {In \cite{li2020energy}, a content offloading/caching scheme in ISTN was proposed .  According to this scheme, the macro base station (MBS) controls the traffic offloading  by allocating the necessary power and assigning the content blocks to the Access Points and to the satellite. This content can be sent to the end users as unicast through the  access points and/or it can be broadcasted through the satellite \cite{li2020energy}. }

In spite of these tremendous efforts in studying traffic offloading in ISTN, we note two main shortcomings in the literature. Firstly, most of the present studies, except \cite{boya}, disregarded that a concrete cooperation between terrestrial and satellite networks during offloading could yield more effective solutions.   {This cooperation  becomes imperative especially because next generation networks will use the terahertz (THz) band   to expand the bandwidth and to satisfy future services  \cite{dang2020what,elayan2020terahertz}. Since, current satellite services use the THz band, future terrestrial networks
are expected to use the same band \cite{elayan2020terahertz}; thus future collaboration schemes between the terrestrial and satellite networks.} Yet, both networks coexist, but they do not cooperate to share task management. 
Secondly, the offloaded traffic is not typically selected, and the traffic requirements are widely overlooked during that.  {This shortcoming becomes more crucial if we take into consideration the heterogeneous use cases and traffic types defined by 5G. Specifically, 5G should address the requirements of the eMBB use case, whose services exchange large payloads over an extended time interval and thus there is a need for high throughput and bandwidth links with moderate reliability levels \cite{tang2019service,bennis2018ultra}. 5G should also answer the requirements of the URLLC use case, whose services have intermittent transmissions between a low number of connected devices and need high reliability and low latency stringently \cite{tang2019service,bennis2018ultra}. To address this issue, several research studied the multiplexing of eMBB services and URLLC services \cite{tang2019service, kassab2019non,anand,alsenwi2019chance,alsenwi2,alsenwi1}. For instance, the authors of \cite{tang2019service} exploited the advantage of orthogonal slicing to multiplex eMBB and URLLC services, where they proposed a framework that maximizes the slice requests' admission and the operator's revenues. The admission of both slice requests depends on the available resources and the required quality-of-service (QoS) such that the eMBB slice uses the multicast transmission and the URLLC slice uses the unicast transmission. A detailed analysis of non-orthogonal multiplexing of eMBB and URLLC services in a fog-radio access network  was conducted in \cite{kassab2019non}. In this regard, the  proposed architecture processed the URLLC traffic in the fog nodes to satisfy its latency constraint. However, the eMBB traffic was processed in the Cloud-RAN to gain high spectral efficiency \cite{kassab2019non}. Various joint eMBB and URLLC scheduler's models were investigated in \cite{anand}, where each scheduler selects  and retrieves some resources from the eMBB traffic and allocates them to the URLLC traffic given its stringent latency constraint by using the superposition/puncturing technique. This technique should satisfy the delays tolerable by the URLLC service and mitigate the rate loss of the eMBB service. The puncturing technique was also used in  \cite{alsenwi2019chance,alsenwi2,alsenwi1} to multiplex the eMBB service and the URLLC service. The authors formulate the resources allocation as 2-Dimensions Hopfield neural network and the resources scheduling as an optimization problem with a chance constraint \cite{alsenwi2019chance}. Unfortunately, all the aforementioned  research overlooks the severe capacity limitation from which terrestrial backhauls are suffering. Moreover, it rules out the dense network scenario since a modest number of bases stations and users was set up.} 

 {   To overcome the previously discussed shortcomings, we propose to manage the multiplexing of the eMBB traffic and the URLLC traffic in ISTN. We advocate that both networks, satellite and terrestrial,  cooperate such that each traffic is precisely steered towards the transport network that meets its requirements. To the best of our knowledge, our paper is the first to  propose a scheme for traffic offloading in ISTN to satisfy heterogeneous traffic requirements. Our main contribution is to consider the offloading of heterogeneous traffic types in ISTN. Specifically, our scheme  offloads the URLLC traffic in the terrestrial backhaul to satisfy its stringent requirement in terms of latency and the eMBB traffic in the satellite backhaul since it needs high data rate but is not very sensitive to delay. Our scheme aims to relieve the constraints on the terrestrial backhaul in the context of dense networks.} To this end, we design the proposed joint ISTN system in two steps; first, we optimize the scheduling of the eMBB traffic and the URLLC traffic by using the  {puncturing} technique, while taking into account the terrestrial backhaul capacity limitation. Second, we adjust the association between the satellite and the small base stations (SBS) while prioritizing URLLC traffic over eMBB traffic. We then evaluate the performance of our scheme in three satellite systems: Telsat LEO, OneWeb, and Starlink. As proved by the results, our scheme reduces the transmission delay of the URLLC packets and decreases the number of dropped eMBB packets; hence it improves the availability of the network. 

 {The remainder of this paper is structured as follows :  In section II,  our system model is introduced. In section III and section IV, our model is formalized in terrestrial networks and in ISTN, respectively. In section V, some performance metrics of our proposed scheme are derived analytically. In section VI, the numerical results are discussed. Finally, the paper is concluded in Section VII.}

\section{System Model}

Our system model consists of a satellite network and a terrestrial network as depicted in Fig. \ref{fig:model}. 
The satellite constellation consists of a set $\mathcal{S}=\{{s}_{1},{s}_{2},{\dots},{s}_{M}\}$ of $M$ LEO satellites. In the scope of this paper, we consider the downlink in the satellite network with a link capacity per beam ${C}_{\mathrm{Sat}}$.  {The used band in the satellite network is the Ka-band \cite{boya}.}  The terrestrial network consists of a set $\mathcal{BS}=\{{bs}_{1},{bs}_{2},{\dots},{bs}_{N}\}$ of $N$ small base stations in small cells $i$ that serve each a set  $\mathcal{U}$ of U UEs. The terrestrial backhaul between the small base station and the macro base station is characterized with a link capacity ${C}_{\mathrm{Ter}}$.  {The used band in the terrestrial network is the C-band \cite{boya}.}

\begin{figure}[htp]
    \centering
    \includegraphics[width=3.5in]{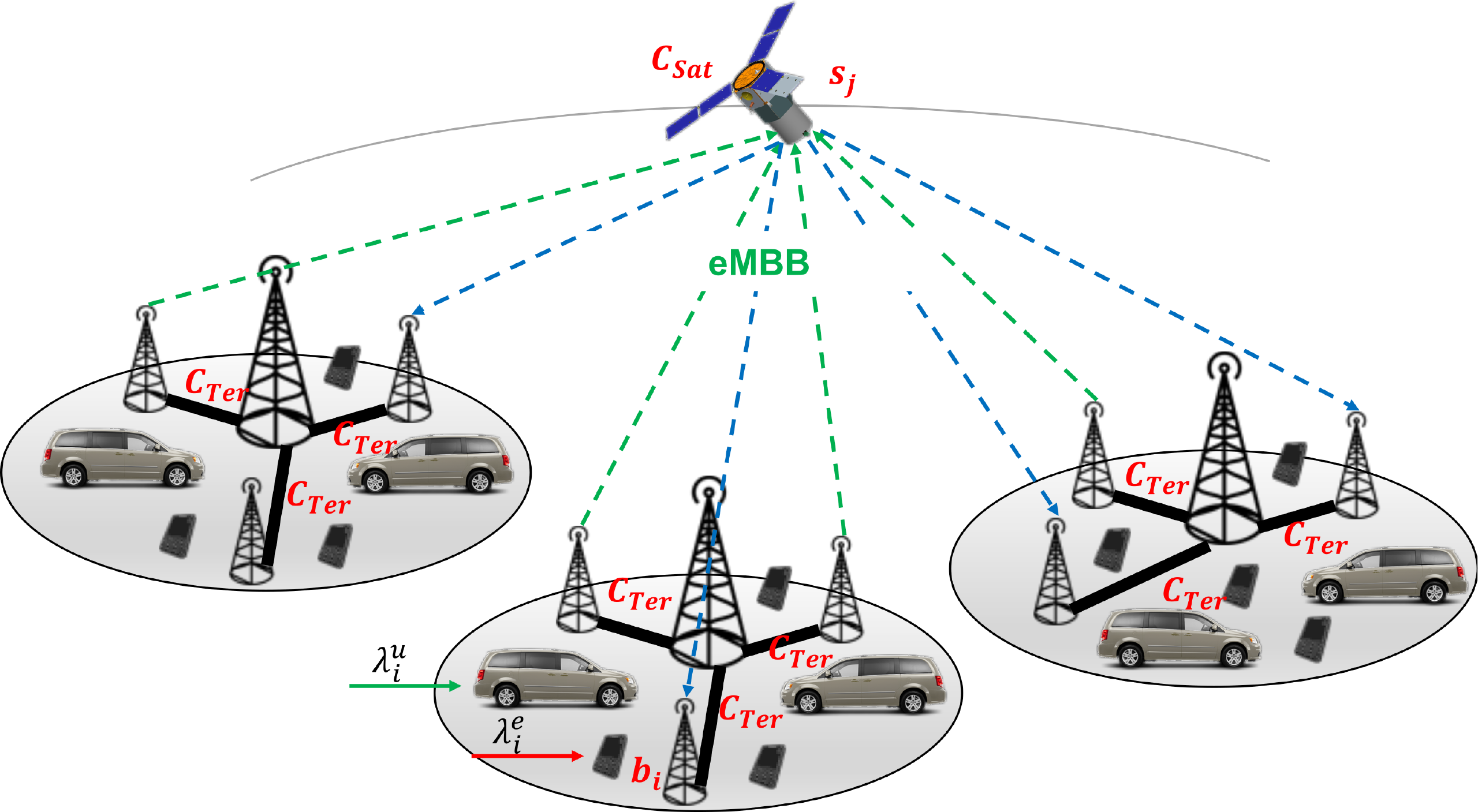}
    \caption{System model of QoS sensitive offloading in ISTN.}
    \label{fig:model}
\end{figure}

\subsection{Traffic Model}

Let us assume that the UE belongs to either eMBB slice or
URLLC slice such that  $U={U}_{\mathrm{e}}{\;\cup\;}{U}_{\mathrm{u}}$ where ${U}_{\mathrm{e}}=\{1,..,{U}_{1}\}$  { is the set of $U_1$ eMBB users} and ${U}_{\mathrm{u}}=\{1,..,{U}_{2}\}$  {is the set  of $U_2$ URLLC users; such that $U_1\geq U_2$.} 
 {The  eMBB traffic   is   generated   by   applications   that   exchange large  payloads  over  an  extended  time  interval \cite{tang2019service}. Therefore, the inter-arrival time can be modeled with an exponential distribution whose events occur continuously and independently. Accordingly, the eMBB flow requests follow a Poisson process with arrival rate  ${\lambda}_{i} ^{\mathrm{e}} $. However, the URLLC traffic  is  generated  by  applications  with  intermittent transmissions that exchange more or less important payload during short periods of time \cite{tang2019service}. Therefore, it can be modeled with a power-law distribution namely a Pareto distribution with arrival rate  ${\lambda}_{i}^{\mathrm{u}} $ and parameters $a$ and ${x}_{m}$ \cite{anand,alsenwi1,alsenwi2019chance,bennis2018ultra,abdel2018ultra,malik2018interference,3GPPURLLC}. Flow sizes are independently and identically distributed} with mean  $\frac{1}{{\mu }_{i}^{\mathrm{e}}}$ for eMBB slice and  $\frac{1}{{\mu }_{i}^{\mathrm{u}}}$ for URLLC slice, such that $\frac{1}{{\mu }_{i}^{\mathrm{e}}}$ and  $\frac{1}{{\mu}_{i}^{\mathrm{u}}}$ represent the packet size for each traffic type, respectively. Given the delay constraints, the small base station  sends the URLLC traffic  in the terrestrial backhaul and offloads the eMBB traffic to the satellite.

\subsection{Base Station Load}
Each small base station   ${bs}_{i}$ is characterized by a specific bandwidth 
${W}_{i}$ and a downlink transmit power  ${P}_{i}$. We assume that the transmit power of the small base station  is the same for all its associated users. The signal-to-noise ratio (SNR) experienced by a user $v$  served by the small base station  ${bs}_{i}$ is given by,
\begin{equation}
\mathrm{SNR}_{iv}=\frac{{P}_{i}\;{g}_{\mathit{i}}^{\mathit{v}}}{{N}_{\mathrm{p}}+\sum
_{i'{\in}\left\{1,..,N\right\}/i}{{P}_{i'}\;{g}_{i'}^{\mathit{v}}}},
\end{equation}
where ${g}_{\mathit{i}^{\mathit{v}}}$ is the channel gain between small base station  ${bs}_{i}$ and user $v$, ${N}_{\mathrm{p}}$ is the noise power of AWGN, and $\sum_{i'{\in}\left\{1,..,N\right\}/i}{{P}_{i'}\;{g}_{i'}^{\mathit{v}}}$ is the interference caused by the other small base stations, where we consider it as negligible in the scope of this paper.  {Indeed, we assume that the proposed system implements the necessary techniques  of inter-cell interference coordination  to avoid or neglect it (e.g. frequency reuse).}

 {We denote the frequency resources allocated to eMBB user $v$  in one slot by ${b}_{v}$ and the punctured frequency resources from eMBB user $v$ to the URLLC traffic due to the hard latency constraints by ${f}_{v}$}.  Thus, the SNR experienced by a URLLC user $v$ is given by \begin{equation}
\mathrm{SNR}_{iv}^{\mathrm{u}}=\frac{{P}_{i}\;{g}_{\mathit{i}}^{\mathit{v}}}{{N}_{0}\;{f}_{v}},
\end{equation}
 where ${N}_{0}$ is the noise power density.

On the other hand, the SNR experienced by an eMBB user $v$ is given by,
\begin{equation}
\mathrm{SNR}_{iv}^{\mathrm{e}}=\frac{{P}_{i}\;{g}_{\mathit{i}}^{\mathit{v}}}{{N}_{0}\;\left({b}_{v}-{f}_{v}\right)}.
\end{equation}

Let  $\gamma:= \frac{{P}_{i}\;{g}_{\mathit{i}}^{\mathit{v}}}{{N}_{0}}$, then the maximum data transmission rate of an URLLC user $v$  is given by \cite{sap2016optimal}, \cite{alsenwi1}:
\begin{equation}
{r}_{\mathit{iv}}^{\mathrm{u}}={f}_{v}\;{\log}_{2}\left(1+\frac{\gamma}{{f}_{v}}\right).
\end{equation}

The maximum data transmission rate of an eMBB user $v$  is given by \cite{sap2016optimal,alsenwi1}:
\begin{equation}
{r}_{\mathit{iv}}^{\mathrm{e}}=\left({b}_{v}-{f}_{v}\right){\log
}_{2}\left(1+\frac{\gamma}{{b}_{v}-{f}_{v}}\right).
\end{equation}

The total load at the $i$-th cell  is given by \cite{sap2016optimal},
\begin{equation}
{L}_{i}={L}_{i}^{\mathrm{u}}+ {L}_{i}^{\mathrm{e}},
\end{equation}
where  ${L}_{i}^{\mathrm{u}}$ and ${L}_{i}^{\mathrm{e}}$ are the load generated by URLLC and eMBB slices, respectively. First, we write ${L}_{i}^{\mathrm{u}}$ as 
\begin{equation}
{L}_{i}^{\mathrm{u}}=\sum_{v=1}^{{U}_{2}}{{r}_{\mathit{iv}}^{\mathrm{u}}}
 =\sum_{v=1}^{{U}_{2}}{{f}_{v}\;{\log}_{2}\left(1+\frac{\gamma}{{f}_{v}}\right)}.
\end{equation}
On the other hand, we write ${L}_{i}^{\mathrm{e}}$ as 
\begin{equation}
 {L}_{i}^{\mathrm{e}}= \sum_{v=1}^{{U}_{1}}{{r}_{\mathit{iv}}^{\mathrm{e}}} 
=\sum_{v=1}^{{U}_{1}}\left({b}_{v}-{f}_{v}\right){\log}_{2}\left(1+\frac{\gamma}{{b}_{v}-{f}_{v}}\right).
\end{equation}

\subsection{Admission Control on URLLC Users}
The capacity limit on the terrestrial link leads us to define an admission control mechanism on URLLC users such that the URLLC demand exceeding the terrestrial system capacity (${C}_{\mathrm{Ter}}$) is blocked \cite{anand}, thus we have
\begin{equation}
\lambda_{i}^{\mathrm{u}}\leq {C}_{\mathrm{Ter}} \; ; \;{\forall\;}i{\;\in}\;\left\{1,..,N\right\}.
\end{equation}
 
\subsection{Downlink Satellite Data Rate}
We assume that each small base station can offload its tasks to one satellite at a time.  {$\eta_{\mathit{ij}}$} denotes the association of the $i-\mathrm{th}$ small base station, i.e.   ${bs}_{i}$,  to the $j-\mathrm{th}$ satellite, i.e., ${s}_{j}$, such that 
 {
\begin{equation}
{\eta}_{\mathit{ij}}=\left\{\begin{matrix} \neq 0 & {bs}_{i}\text{\:can\:  }\text{offload\:  }\text{tasks\:  to\:  } {s}_{j}\\
0 &
\text{otherwise}\end{matrix}\right. .
\end{equation}}

The small base station-satellite association rule can be formulated as 
 {
\begin{equation}
\left\{\begin{matrix}{\eta}_{\mathit{ij}}{\;\in\;}[0,1],&{\forall\;}i{\;\in\;}\left\{1,..,N\right\},j{\;\in\;}\left\{1,..,M\right\} \\ \\ \sum
_{j=1}^{M}{{\eta}_{\mathit{ij}}=1},&{\forall\;}i{\;\in\;}\left\{1,..,N\right\}\end{matrix}\right.
\end{equation}}
where the backhaul link between small base station   ${bs}_{i}$ and satellite  ${s}_{j}$
is characterized by a maximum achievable data rate  ${R}_{\mathit{ij}}$  given by \cite{boya}
\begin{equation}
{R}_{\mathit{ij}}={W}_{j}\;{\log
}_{2}\left(1+\frac{C}{{N}_{0}\;{W}_{j}}\right),
\end{equation}
where ${W}_{j}$ is the bandwidth dedicated to each satellite and shared by its associated small base stations and  $\frac{C}{{N}_{0}}$ is the carrier signal power to noise power density of the satellite.  {
We should note that 
${R}_{\mathit{ij}} \leq C_{\mathrm{Sat}}$; where $\frac{C}{{N}_{0}}$ and $C_{\mathrm{Sat}}$  are fixed by the satellite manufacturer.
}

In the rest of the paper, we focus on a single satellite of the network and we denote the data rate between the fixed satellite and the small base station   ${bs}_{i}$ by  ${R}_{\mathit{j}}$, and  the offloaded amount of eMBB traffic between from the satellite and the $i-\mathrm{th}$ small base station, i.e.,    ${bs}_{i}$ by $\beta_i$.

\section{Resource Allocation in the Terrestrial Networks}
We start first by scheduling the eMBB traffic and the URLLC traffic using the  {puncturing} technique which superimposes the URLLC traffic upon its arrival on the ongoing eMBB transmission \cite{anand,alsenwi2}. 
 {Our aim is to characterize the optimal resource allocation} for URLLC users while maximizing the data rate of eMBB users and taking into account the limited backhaul capacity. 
 {Thus, we formulate the resource allocation problem as follows }
\begin{subequations}
\label{eq:8}
\begin{align}
 & \max_{{f}_{v}} \;\sum_{v=1}^{{U}_{1}}{\left({b}_{v}-{f}_{v}\right){\log
}_{2}\left(1+\frac{\gamma}{{b}_{v}-{f}_{v}}\right)} \\
 \mathrm{s. t.}\quad & \Pr \left(  \sum_{v=1}^{{U}_{2}}{{f}_{v}\;{\log
}_{2}\left(1+\frac{\gamma}{{f}_{v}}\right)}  < \lambda_{i}^{u}   \right)\leq \epsilon \label{eq:8b}\\ 
 & {L}_{i}^{\mathrm{u}}+ {L}_{i}^{\mathrm{e}}\leq{C}_{\mathrm{Ter}}\quad \forall i\; \in \left\{1,..,N\right\}  \label{eq:c}\\
 &\sum_{v=1}^{{U}_{2}}{f}_{v} \leq {W}_{i}^{\mathrm{bs}} \label{eq:13d},
\end{align}
\end{subequations}
where the constraint (\ref{eq:8b}) expresses an acceptable outage probability of URLLC traffic, where $\epsilon$ is a small positive constant, $\epsilon \ll 1$, which guarantees low URLLC outage probability below a negligible threshold.  {Thus, the constraint (\ref{eq:8b}) guarantees that the dropped URLLC packets during the admission control are below a negligible outage probability threshold $\epsilon$ to ensure a URLLC packets transmission reliability.} The constraint (\ref{eq:c}) guarantees that the accepted eMBB load and URLLC load are limited by the terrestrial link capacity, i.e., $C_{\mathrm{Ter}}$. It is worthy emphasizing that extra eMBB packets are dropped in case of overload.  {The constraint (\ref{eq:13d}) guarantees that the sum of the allocated frequency resources to the URLLC users is less than the total bandwidth $W_i^{\mathrm{bs}}$ of the small base station ${bs}_{i}$.}

  We can simplify constraint (\ref{eq:8b}) using the cumulative distribution function (CDF) of $\lambda_{i}^{\mathrm{u}}$, $F_D\left(.\right)$, as follows
\begin{equation}
 \sum_{v=1}^{{U}_{2}}{{f}_{v}\;{\log
}_{2}\left(1+\frac{\gamma}{{f}_{v}}\right)} \geq {F}_{D}^{-1}(1-\epsilon)\;  {.}  \label{eq:14}
\end{equation}
 Since $\lambda_{i}^{\mathrm{u}}$ follows a Pareto distribution with parameters $a$ and ${x}_{m}$, we express (\ref{eq:14}) as
\begin{equation}
 {\left(\sum_{v=1}^{{U}_{2}}{{f}_{v}\;{\log
}_{2}\left(1+\frac{\gamma}{{f}_{v}}\right)}\right)}^{a} \geq \frac{{{x}_{m}}^{a}}{\epsilon}\;  {,}
\end{equation}
 where $a$ is a constant that determines the distribution shape and  $x_m$ is also a positive scale distribution parameter. In this work, we set $x_m=1$ because we assume that the URLLC traffic exhibits self-similarity over time scales of one second and above \cite{crovella1997self}. In addition, we set $a=1$ to reflect a wider variability of the URLLC traffic in the duration of packet generation (ON period) and inter-packets generation (OFF period), which enables us to write (\ref{eq:8b}) in a convex form.  
 However, the constraint (\ref{eq:c}) is a non-convex function in $f_v$, which makes the problem (\ref{eq:21}) non-concave.  To deal with this problem, we propose expressing \eqref{eq:c} as a difference of convex (DC) functions. Therefore, we decompose \eqref{eq:c} as the difference between two convex functions $\mathcal{F}$ and  $\mathcal{G}$ 
\begin{equation}
\mathcal{F}(f_v)-\mathcal{G}(f_v) \leq{C}_{\mathrm{Ter}} \label{eq:d}\; {,}
\end{equation}
such that
 {
\begin{equation}
    \mathcal{F}(f_v) =\sum_{v=1}^{{U}_{2}}{{f}_{v}\;{\log}_{2}\left({f}_{v}+{\gamma}\right)}+\sum_{v=1}^{{U}_{1}}{\left({b}_{v}-{f}_{v}\right){\log}_{2}\left({b}_{v}-{f}_{v}+\gamma\right)}
\end{equation}
and \begin{equation}
    \mathcal{G}(f_v)=\sum_{v=1}^{{U}_{2}}{{f}_{v}\;{\log}_{2}\left({f}_{v}\right)}+\sum_{v=1}^{{U}_{1}}{\left({b}_{v}-{f}_{v}\right){\log}_{2}\left({b}_{v}-{f}_{v}\right)}. 
\end{equation}
First, we prove that $\mathcal{F}$ and  $\mathcal{G}$ are convex.
Indeed, $\mathcal{F}$ is convex since 
\begin{equation}\mathcal{F''}(f_v)=\sum_{v=1}^{{U}_{2}}{\frac{2\gamma+f_v}{(\gamma+f_v)^2}}+\sum_{v=1}^{{U}_{1}}{\frac{2b_v+\gamma-f_v}{(\gamma+b_v-f_v)^2}} \geq 0   \end{equation} 
$ \forall \; f_v \in \left[ 0,b_v \right], \;  \mathcal{G}$ is also convex since $\mathcal{G}(f_v)=\mathcal{F}(f_v)|_{\gamma=0}$.}

 {
Then, we approximate $\mathcal{G}(f_v)$ using the first order representation of Taylor series at $f'_{v}$, which gives the following lower bound as follows \cite{boyd2004convex},
 \begin{equation} \label{G_taylor}
\mathcal{G}(f_v) \leq \mathcal{G}(f'_{v})+ \frac{\partial \mathcal{G}}{\partial f_v}\left(f'_{v}\right)(f_{v}-f'_{v}),
\end{equation}}
where $\frac{\partial \mathcal{G}}{\partial f_v}$ is the  first  derivative  of $\mathcal{G}(f_v)$ with  respect to $f_{v}$, and is written as 
 {
\begin{equation}
\frac{\partial \mathcal{G}}{\partial f_v}=\sum_{v=1}^{{U}_{2}}{\left(1+{\log}_{2}\left({f}_{v}\right)\right)}-\sum_{v=1}^{{U}_{1}}{\left(1+{\log}_{2}\left(b_v-{f}_{v}\right)\right)}.
\end{equation}
Thus, we can write \eqref{G_taylor} as follows 
\begin{equation}
\mathcal{G}(f_v) \leq \sum_{v=1}^{{U}_{2}}{\left(1+{\log}_{2}\left({f}_{v}^{\prime}\right)\right)}-\sum_{v=1}^{{U}_{1}}{\left(1+{\log}_{2}\left(b_v-{f}_{v}^{\prime}\right)\right)}+ \left[\sum_{v=1}^{{U}_{2}}{\left(1+{\log}_{2}\left({f}_{v}^{\prime}\right)\right)}-\sum_{v=1}^{{U}_{1}}{\left(1+{\log}_{2}\left(b_v-{f}_{v}^{\prime}\right)\right)}\right](f_{v}-f'_{v}).\label{eq:22}
\end{equation}
}

Therefore, a simplified approximated problem of (\ref{eq:8}) can be expressed by using the aforementioned first-order approximation (\ref{eq:22}),  { which yields an upper bound for the left-hand-side (LHS) of the original constraint \eqref{eq:c}. Thus. we can rewrite the resource allocation as follows,}
\begingroup
\begin{subequations}
\label{eq:21}
\begin{align}
 & \max_{{f}_{v}} \;\sum_{v=1}^{{U}_{1}}{\left({b}_{v}-{f}_{v}\right){\log
}_{2}\left(1+\frac{\gamma}{{b}_{v}-{f}_{v}}\right)} \\
 \mathrm{s.t.}\; &  -\sum_{v=1}^{{U}_{2}}{{f}_{v}\;{\log
}_{2}\left(1+\frac{\gamma}{{f}_{v}}\right)} \leq -\frac{{{x}_{m}}}{\epsilon}\\ 
&  {{\sum_{v=1}^{{U}_{2}}{{f}_{v}\;{\log}_{2}\left({f}_{v}+{\gamma}\right)}+\sum_{v=1}^{{U}_{1}}{\left({b}_{v}-{f}_{v}\right){\log}_{2}\left({b}_{v}-{f}_{v}+\gamma\right)}\;-}}\nonumber\\ 
& {{\left[\sum_{v=1}^{{U}_{2}}{\left(1+{\log}_{2}\left({f}_{v}^{\prime}\right)\right)}-\sum_{v=1}^{{U}_{1}}{\left(1+{\log}_{2}\left(b_v-{f}_{v}^{\prime}\right)\right)}\right](f_{v}-f'_{v})-}}\nonumber\\
& {{\sum_{v=1}^{{U}_{2}}{\left(1+{\log}_{2}\left({f}_{v}^{\prime}\right)\right)}+\sum_{v=1}^{{U}_{1}}{\left(1+{\log}_{2}\left(b_v-{f}_{v}^{\prime}\right)\right)}\;\leq\;{C}_{\mathrm{Ter}}}  \label{eq:23c}}\\
 &\sum_{v=1}^{{U}_{2}}{f}_{v} \leq {W}_{i}^{\mathrm{bs}}.
\end{align}
\end{subequations}
\endgroup

 {
Thanks to formulating the constraint (\ref{eq:c}) in terms of a DC function, and approximating it using first-order Taylor series obtaining \eqref{eq:23c}, the reduced problem \eqref{eq:21} becomes a convex problem. It is worth to emphasize that such an approximation does not violate the problem; however, the proposed bound reduces the feasibility space resulting in a sub-optimal solution. }
 {To solve the non-convex problem \eqref{eq:8}, we use the successive convex approximation (SCA) and solve the approximated convex problem \eqref{eq:21} iteratively where $f'_{v}$ is updates at each iteration using the previous solution of \eqref{eq:21}.}   
\vspace{0.5cm}

{\centering
\begin{minipage}{.7\linewidth}
\begin{algorithm}[H]
 {
\SetAlgoLined
\SetKwInput{KwInput}{Input}
\SetKw{KwInitialize}{Initialize:}
\KwInput{$[{b}_{v}]_{v \;\in \;\{1 \dots {U}_1\}}$,$\;\epsilon$,
$C_{\mathrm{Ter}}$,$\;\gamma$}
\KwInitialize{ $\epsilon_{\mathrm{error}}$,  ${f}^{'}_{v};\; \forall\;{v \:\in \:\{1  \dots {U}_1\}}$} \\
 \While{$\left(|{f}^{'}_{v} - {f}^{*}_{v}|_{v \:\in \:\{1  \dots {U}_1\}}\geq \epsilon_{\mathrm{error}}\right)$}{
Solve problem (\ref{eq:21}) to get $[{f}_{v}]^{*}_{v \:\in \:\{1  \dots {U}_1\}}$ ;\\
 ${f}^{'}_{v}\; \gets  {f}^{*}_{v}\;;\; \forall\;{v \:\in \:\{1  \dots {U}_1\}}$;}}
 \caption{Resource Allocation in the Terrestrial Network}
\end{algorithm} 
\end{minipage}
\par
}
\vspace{0.5cm}

Algorithm 1 solves the resource allocation problem of the terrestrial network using the SCA approach. First, we start with an initial arbitrarily values for  $[{f'}_{v}]_{v \;\in\;\{1..{U}_1\}}$ and stopping error, $\epsilon_{\mathrm{error}}$, for the iterative algorithm. We choose to start with 
small values of  $f^{'}_{v}$, for instance $f^{'}_{v}=0.1{b}_{v}$,  since our objective function aims to take as little as possible from the eMBB users resources ${b}_{v}$. In each iteration of Algorithm 1, we solve the convex optimization problem (\ref{eq:21}) using the Interior Point algorithm, where ${f'}_{v}$ uses the solution of the previous iteration.  

Once the problem (\ref{eq:8}) is solved, we obtain the optimal scheduling of eMBB users and URLLC users in the terrestrial network. We consider the optimized solution $[{f}^{*}_{v}]_{v \;\in \;\{1..{U}_1\}}$ in order to estimate the loads $[{L}_{i}^{\mathrm{e}}]^{*}_{\;i\;\in\;\{1,..N\}}$ and $[{L}_{i}^{\mathrm{u}}]^{*}_{\;i\;\in\;\{1,..N\}}$ of all small base stations. Using these parameters, we model the  inter-working of the satellite with the terrestrial network in the next section. 
\section{Resource Allocation in  the Integrated Satellite Terrestrial Network}
We use the satellite networks that present additional backhaul links for the terrestrial users  to fulfill the QoS requirements of user slices (i.e., eMBB and URLLC).  { We aim to adjust the association between the satellite and $N$ small base stations by determining $({\alpha},{\beta})$ where $\alpha={[\alpha_1,..,\alpha_i,. .,\alpha_N]}^T$  denotes the bandwidth allocation vector and $\beta={[\beta_1,..,\beta_i,..,\beta_N]}^T$ denotes the offloading vector. Specifically,  $\alpha_{i\in \{1,..,N\}}$ is the bandwidth proportion allocated by the satellite to the small base station ${bs}_{i\in \{1,..,N\}}$ and ${\beta}_{i\in \{1,..,N\}}$ is the proportion of the offloaded eMBB traffic ${L}_{i}^{e*}$ from the small base station ${bs}_{i\in \{1,..,N\}}$ to the satellite.  Our scheme takes into account the limitations on backhaul capacities on both terrestrial and satellite networks.}

Let $F_{i}={\beta}_{\mathit{i}}\;{L}_{i}^{\mathit{e*}}$ denotes the offloaded traffic to the satellite by the small base station ${bs}_{i}$. Our problem can be posed as follows:

\begin{subequations}
\label{eq:20}
\begin{align}
&\max_{\alpha,\beta} \quad [F_{1},F_{2},..,F_{i},..,F_{N}] \;\;\;\;\;  \quad{\forall\;}i{\;\in}\left\{1,..,N\right\} \label{eq:20a} \\
&\mathrm{s.t.} \nonumber \\
&{F_{i}\leq {\alpha}_{i}\; W\;{\log
}_{2}\left(1+\frac{C}{{\alpha}_{i}\;W\;{N}_{0}}\right) \quad \quad \; \; {\forall\;}i{\;\in}\left\{1,..,N\right\}}\label{eq:20b}\\
  &{\sum_{i=1}^{N}{{F}_{i}} \leq C_{\mathrm{Sat}} \label{eq:20c}}\\
   & { {\alpha}_{i} - {\alpha}_{i'} \leq \frac{1}{ \delta({L}_{i}^{\mathrm{u}},{L}_{i'}^{\mathrm{u}})}-1\quad\qquad\qquad{\forall\;}i,i'{\in}\left\{1,..,N\right\}} \label{eq:i}\\[4pt]
 &{ {\beta}_{\mathit{i}}-{\beta}_{\mathit{i'}}\leq \frac{1}{ \delta({L}_{i}^{\mathrm{u}},{L}_{i'}^{\mathrm{u}})}-1\quad\qquad\qquad{\forall\;}i,i'{\in}\left\{1,..,N\right\}} \label{eq:j}\\[4pt]
  &{ 0\leq {\alpha}_{i} \leq 1   \qquad \qquad\qquad\qquad \qquad  \qquad \quad {\forall\;}i{\;\in}\left\{1,..,N\right\}}\\[4pt]
 &{\sum_{i=1}^{N}{{\alpha}_{i}}=1}\\[4pt]
 &{ 0\leq {\beta}_{\mathit{i}} \leq 1 \qquad \qquad \qquad \qquad \qquad  \qquad \quad {\forall\;}i{\;\in}\left\{1,..,N\right\}}\label{eq:20h},
\end{align}
\end{subequations}
where \eqref{eq:20} represent a maximization problem for multi-objective functions with several constraints \eqref{eq:20b} -- \eqref{eq:20h}. First, the constraint \eqref{eq:20b}  {ensures that the eMBB traffic offloaded by the base station ${bs}_i$  does not exceed the resources $\alpha_i$  allocated by the satellite }.  The constraint (\ref{eq:20c}) reflects the limitation on the satellite backhaul capacity, where the sum of eMBB loads should be less than $C_{\mathrm{Sat}}$. However, the constraints (\ref{eq:i}) and (\ref{eq:j}) are added to prioritize the small base station that has a more important URLLC traffic to offload its eMBB traffic to the satellite backhaul where
\begin{equation}
    \delta({L}_{i}^{\mathit{u}},{L}_{i'}^{\mathit{u}})=\left\{\begin{matrix} 1 \; \text{if} \; {L}_{i}^{\mathit{u}}\leq {L}_{i'}^{\mathit{u}}\\0, \;
\text{otherwise\;.}\end{matrix}\right..
\end{equation}

To  solve  our multi-objective optimization, we  use  the weighted sum method, where each  objective function ${F}_{i \in \{1,..,N\}}$ is assigned a weight ${\omega}_{\mathit{i}\in \{1,..,N\}}$. One way to fix the weights ${\omega}_{\mathit{i}\in \{1,..,N\}}$ in our optimization problem (\ref{eq:20}) is based on the location of the small base station ${bs}_{{\mathit{i}\in \{1,..,N\}}}$ within the coverage area of the satellite. The preponderant weights are allocated to the small base stations located at the edge of the coverage area to boost their offloading. These weights can be periodically changed between the $N$ small base stations based on the visibility window of the satellite. Our problem (\ref{eq:20}) can be restated as follows
\begin{subequations}
\begin{align}
&\max_{\alpha,\beta}\quad \sum_{i=1}^{N}{{\omega}_{\mathit{i}}\;{\beta}_{\mathit{i}}\;{L}_{i}^{\mathrm{e}*}} \nonumber\\
&\text{s.t.}\nonumber\\
& {{\beta}_{\mathit{i}}\;{L}_{i}^{\mathrm{e}*}\leq {\alpha}_{i}\; {W}_{j}\;{\log
}_{2}\left(1+\frac{{P}_{j}\;{g}_{\mathit{ij}}}{{N}_{0}\;{\alpha}_{i}\;{W}_{j}}\right) \;  {\forall\;}i{\;\in}\left\{1,..,N\right\}} \tag{\ref{eq:20b}} \\[3pt]
&{\sum_{i=1}^{N}{{\beta}_{\mathit{i}}\;{L}_{i}^{\mathrm{e}*}} \leq C_{\mathrm{Sat}}} \tag{\ref{eq:20c}}\\[5pt]
& \eqref{eq:i}-\eqref{eq:20h}.\nonumber
\end{align}
\end{subequations}
 Our problem (\ref{eq:20}) is convex with a linear objective function, a convex constraint (\ref{eq:20b}) and linear constraints (\ref{eq:20c})-(\ref{eq:20h}). Thus we can solve by any convex optimization tool obtaining the optimal proportion $\beta_{i\;\in\;\{1,..N\}}$ of the eMBB traffic that should be offloaded by each small base station to the satellite and the optimal proportion $\alpha_{i\;\in\;\{1,..N\}}$ of the satellite bandwidth that should be allocated to each small base station according to the offloaded traffic. By using the optimal solution ${(\alpha_{i},\beta_{i})}_{i\;\in\;\{1,..N\}}$, the eMBB load  $L_i^{\mathrm{eT}}$ that should be transmitted by the small base station ${bs}_{i\;\in\;\{1,..N\}}$ through the terrestrial backhaul  can be computed  in ISTN as
\begin{equation}
   L_i^{\mathrm{eT}}= (1-\beta_i)L_i^{\mathrm{e}*}. \label{eq:23}
\end{equation}
This load, $L_i^{\mathrm{eT}}$, is fundamental to estimate the URLLC delay and the dropped eMBB traffic experienced in the terrestrial backaul. Accordingly, we can compare the performance of our scheme operating  in  ISTN to a benchmark scheme operating in the terrestrial network as follows in the next section. 
\section{Terrestrial Delay Analysis}
 In this section, we study the  URLLC delay experienced in the terrestrial backaul analytically. This delay depends not only on the URLLC traffic ${L_i^\mathrm{u}}^*$of the small base station ${bs}_{i\;\in\;\{1,..N\}}$, but also on the eMBB traffic sent through this backhaul. This latter traffic is given by equation (\ref{eq:23}); where $\beta_{i}$ is the optimal solution of our problem (\ref{eq:20}) in the case of ISTN. In the case of terrestrial network, $\beta_{i}=0$.    
 
In this analysis, the terrestrial link of the small base station ${bs}_i$ is modeled as M/G/1 queue where :
\begin{itemize}
    \item the arrival rate of the eMBB packets and URLLC packets $\lambda_{i}$ to  small base station ${bs}_i$ follows a Poisson process  given by $\lambda_{i}=\lambda_{i}^{\mathrm{e}}+\lambda_{i}^{\mathrm{u}}$
    \item the service time distribution is general. 
\end{itemize}
\subsection{Mean Waiting Time }
We assume that the scheduling discipline is First Come First Serve (FCFS). The waiting time $\mathcal{T}_k$ of packet $k$ scheduled after $K$ packets  is  given by:

\begin{equation}
    \mathcal{T}_k=\mathcal{R}_k+\sum_{n=k-K}^{k-1}{\mathcal{X}_n}\; ,
\end{equation}
where   $\mathcal{X}_n$ is the service time of packet $n$ that arrived before packet $k$ and $\mathcal{R}_k$ is the residual service time.

Using Little's formula, we obtain the average waiting time $ \overline{\mathcal{T}}_k$ as :
\begin{equation}
\overline{\mathcal{T}}_k=\frac{\overline{\mathcal{R}}_k}{1-\rho_i}\; ,
\end{equation}
where $\rho_i$ is the total load in the small cell $i$ 
\begin{equation}
    \rho_i=\rho_{i}^{\mathrm{e}}+\rho_{i}^{\mathrm{u}}=\frac{\lambda_{i}^{\mathrm{e}}}{\mu_\mathrm{e}}+\frac{\lambda_{i}^{\mathrm{u}}}{\mu_\mathrm{u}}\;.
\end{equation}
 
$\rho_i$ is expressed in both networks as follows
\begingroup
\normalsize
\begin{equation}
\rho_i=\left\{\begin{matrix}  \frac{L_{i}^{\mathrm{e}*}+L_{i}^{\mathrm{u}*}}{C_{\mathrm{Ter}}} & \text{in the terrestrial network}\\[7pt]
\frac{(1-\beta_i)L_{i}^{\mathrm{e}*}+L_{i}^{\mathrm{u}*}}{C_{\mathrm{Ter}}} &
\text{in ISTN}\end{matrix}\right.
\end{equation}
\endgroup

The first moment of the residual service time can be developed as :
\begin{equation}
\overline{\mathcal{R}}_k=\frac{\rho_{i}}{2}\;\frac{\overline{\mathcal{X}^2}}{\overline{\mathcal{X}}}\;. \label{eq:26}
\end{equation}

In order to evaluate the mean service time $\overline{\mathcal{X}}$, we approximate our general distribution to an exponential distribution with a mean service rate ${\mu_e}$. Therefore, the first moment of the residual service time can be simplified, based on (\ref{eq:26}), as   
\begin{equation}
\overline{\mathcal{R}}_k=\frac{\rho_{i}}{2}\;\frac{\frac{2}{\mu_\mathrm{e}^2}}{\frac{1}{\mu_\mathrm{e}}}=\frac{\rho_{i}}{\mu_\mathrm{e}}\;.
\end{equation}

\subsection{Distribution of the Waiting Time }

To find the distributions of the time spent in the system and in the queue, we consider the Laplace transform  approach. Let $T$ be the total time spent in the system, $T_q$  the total waiting time before the service begins and $X$ the service time. This yields that $T=Q+X $ such that $Q$ and $X$ are independent. 

We denote with $L_T$, $L_Q$ and $L_B$ the Laplace transforms of $T$, $T_q$ and $X$ respectively.

\begin{equation*}
    L_T(s)=E[e^{-s\;T}]=E[e^{-s\;(Q+X)}]=E[e^{-s\;Q}]\;.\;E[e^{-s\;X}]=L_Q(s)\;.\;L_B(s)\;.
\end{equation*}
            
            Hence 
            \begin{equation}
                L_Q(s)=\frac{L_T(s)}{L_B(s)} \label{eq:28}
            \end{equation}
Based on equation (5.100) in \cite{kleinrock1976queueing}, 
 \begin{equation}
L_T(s)= L_B(s)\; \frac{s\;(1-\rho_i)}{s-\lambda_i+\lambda_i\;L_B(s)}\;.\label{eq:29}
 \end{equation}

Substituting (\ref{eq:29}) in  (\ref{eq:28}) yields
 \begin{equation}
                L_Q(s)=\frac{s\;(1-\rho_i)}{s-\lambda_i+\lambda_i\;L_B(s)}\;.
            \end{equation}

In order to evaluate $L_Q$, we approximate our general distribution to an exponential distribution with a mean service rate ${\mu_\mathrm{e}}$. Accordingly, 
\begin{equation}
                L_B(s)=\frac{\mu_\mathrm{e}}{\mu_\mathrm{e}+s}\;.
            \end{equation}
Therefore

\begin{equation}
 L_Q(s)=(1-\rho_i)\;(1+\frac{\lambda_i}{s+\mu_\mathrm{e}-\lambda_i})\;,\label{eq:32}
            \end{equation}

and 
\begin{equation}
L_T(s)= \frac{\mu_e-\lambda_i}{s+\mu_\mathrm{e}-\lambda_i}\;.\label{eq:33}
 \end{equation}
 
 Let $f_Q$ and $f_T$ denote the Probability Distribution Functions (PDF) of  $Q$ and $T$ respectively. To determine the waiting time distributions, we apply the inverse Laplace transform on equations (\ref{eq:32}) and (\ref{eq:33}). This leads to

 \begin{align}
&f_Q(t)=(1-\rho_i)\;\left(\delta(t)+e^{(\lambda_i-\mu_\mathrm{e})t}\right)\;, \label{eq:34}\\
& \text{where}\quad \delta(t) =
    \begin{cases} \nonumber
            \infty, &         \text{if } t=0\;,\\
            0, &         \text{otherwise}.
    \end{cases}
 \end{align}
 
 and 
\begin{equation} 
 f_T(t)=(\mu_\mathrm{e}-\lambda_i)\;e^{-(\mu_\mathrm{e}-\lambda_i)t} \;.\label{eq:35}
  \end{equation}
  
  The CDF of the total waiting time $F_T$ can finally be obtained  as  
  
  \begin{equation} 
 F_T(x)=\int^x_{-\infty}{f_T(t)\;dt}=1-e^{-(\mu_\mathrm{e}-\lambda_i)x},\;x\geq0\;. \label{eq:36}
  \end{equation}
  
\section{Results and Discussion}
In this section, we evaluate our scheme performance in terms of latency and availability. In this regard, we assess the variation of the URLLC delay in the terrestrial backhaul and  the dropped volume of eMBB traffic.  We conducted our simulations based on the features of three LEO constellations. The simulations parameters are detailed in Table \ref{table:1} for the terrestrial network and in Table \ref{table:2} for the satellite network.
\begin{table}
\caption{ Terrestrial Network Parameters}
\centering
\begin{tabular}{|c |c| } 
 \hline
Parameter & Numerical Value\\ [1ex] 
 \hline\hline 
 Small cell radius (m) & 500 \\[1ex] 
 Bandwidth (Mhz) & 100 \\[1ex] 
 RB Bandwidth (Mhz) & 0.18 \\[1ex]
 $N_o$ (dbm/Hz) & -174 \\ [1ex] 
 SBS transmission power (W) & 20 \\ [1ex]
 URLLC packet size (B) \cite{ETSITR138913} & 30 \\[1ex]
 eMBB packet size (B) \cite{ETSITR138913} & 100 \\[1ex]
 \hline 
\end{tabular}
\label{table:1}
\end{table}
 
\begin{table}[h]
\caption{ Satellite Features per Downlink Beam \cite{del2019technical}}
\centering
\begin{tabular}{|c |c| c |c|} 
 \hline
Parameter & Telsat & OneWeb & SpaceX \\ [1ex] 
 \hline\hline 
 Bandwidth (Ghz) & 0.25 & 0.25 & 0.25 \\ [1ex] 
 Satellite Capacity (Mbps) & 558.7 & 599.4 & 674.3 \\[1ex] 
  $Rx\frac{C}{N_0}$ (db) & 9.6 & 10.5 & 12 \\
 [1ex] 
 \hline 
\end{tabular}
\label{table:2}
\end{table}

 { First, we prove the convergence of our proposed algorithm for resource allocation in the terrestrial network.  Therefore, we plot the eMBB sum rate versus the number of iterations for different scenarios of terrestrial backhaul capacities as shown in Fig.\ref{fig:iter}.  We observe a gradual increase of the eMBB sum rate with increasing the number of iterations. Then, the eMBB sum rate saturates at a maximum value within 70 iterations. }
\begin{figure}[H]
    \centering
    \includegraphics[width=3.5in]{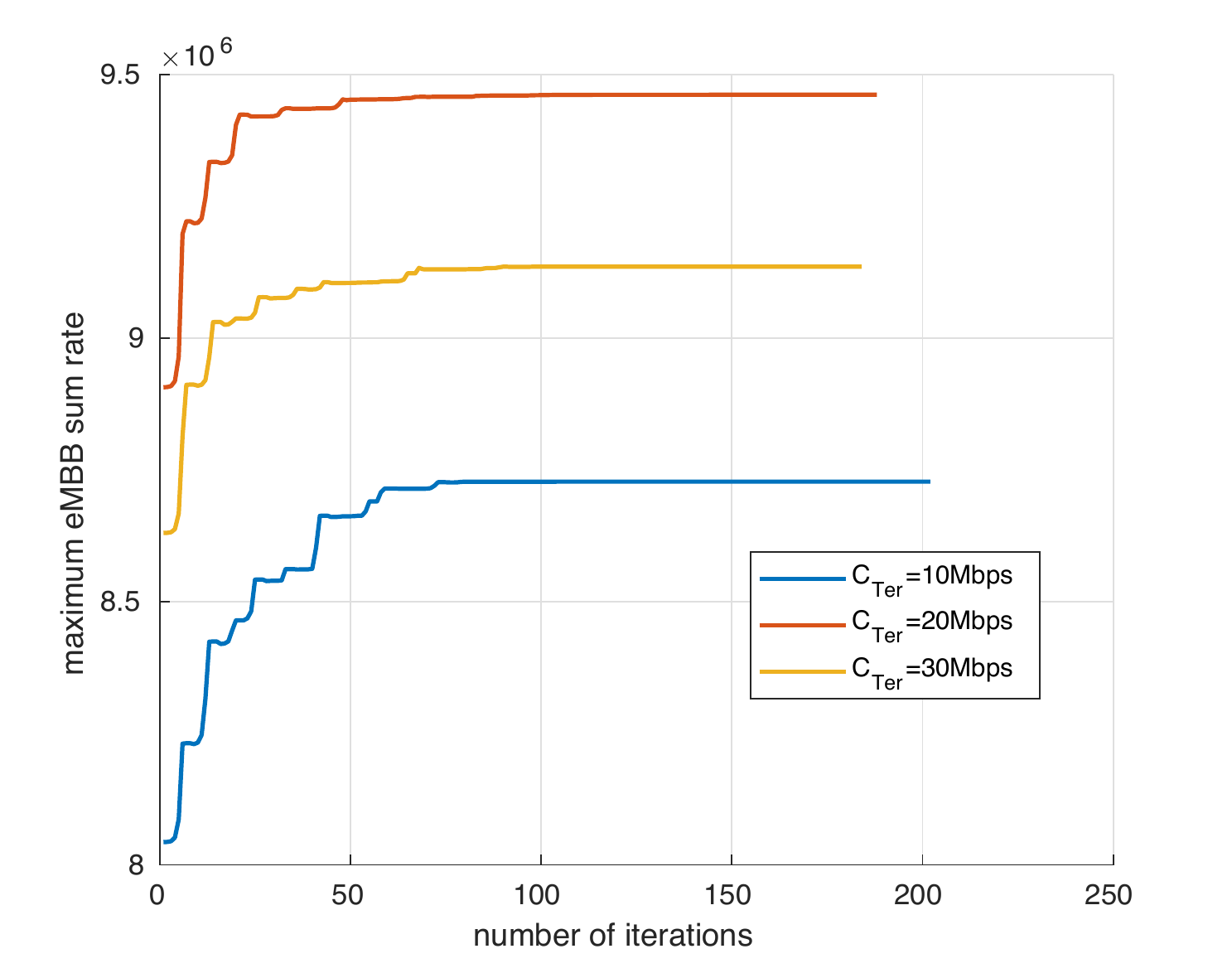}
    \caption{Sum rate vs number of iterations}
    \label{fig:iter}
\end{figure}

Then, we evaluate the performance of our scheme according to the variation of the link capacity and of the link load respectively. We compare our results obtained in ISTN to a benchmark that consists of a network with a terrestrial backhaul only. To fairly compare both schemes, the considered benchmark has an equivalent terrestrial backhaul capacity $C_{\mathrm{Ter}}^{'}$ such that 
\begin{equation}
C_{\mathrm{Ter}}^{'}=C_{\mathrm{Ter}}+\frac{C_{\mathrm{Sat}}}{N}\;.
\end{equation}

In  the  first  simulation, we investigate the role of satellites integration in terrestrial networks to alleviate the capacity limitation in  dense networks.  We  vary  the  link capacity ${C}_{\mathrm{Ter}}$ between 10 Mbps and 100 Mbps for a fixed network load $\rho = 0.8$ that characterizes highly loaded networks. We study first the variation of the URLLC delay experienced in the terrestrial backhaul in both networks.

\begin{figure}[h]
    \centering
    \includegraphics[width=3.5in]{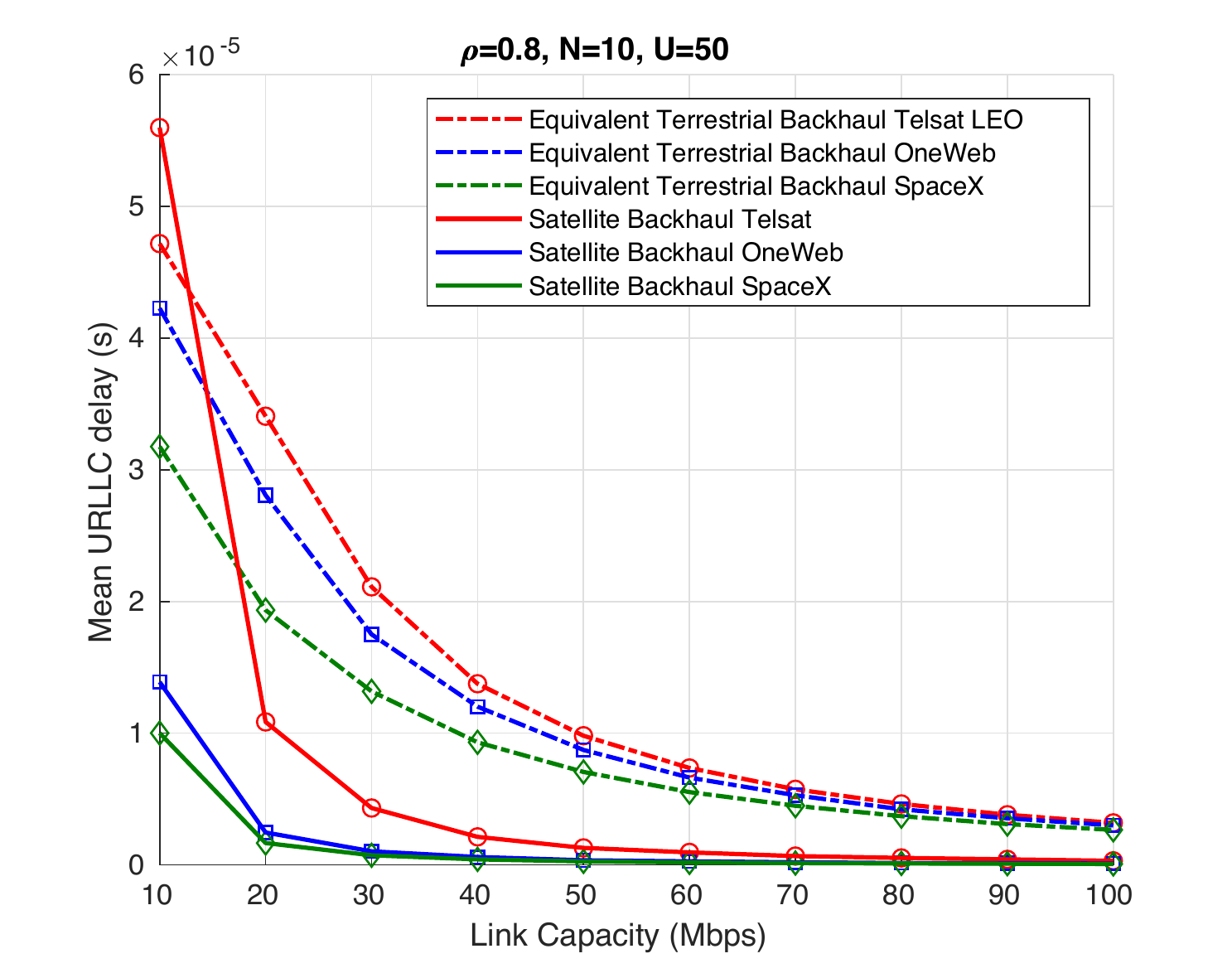}
    \caption{Mean URLLC Delay as function of Link Capacity.}
    \label{fig:4}
\end{figure}

Based on the results depicted in Fig.\ref{fig:4}, we observe that the delay decreases as expected with larger capacities. We notice also that our scheme, based on satellite and terrestrial network integration, achieves lower delays than the terrestrial benchmark. Interestingly, our scheme outperforms significantly the  benchmark for the small link capacities (i.e. less than 30 Mbps). This finding highlights the key role that our scheme plays in dense networks where link capacities are considerably limited. 
 
Then, we study the mean amount of the dropped eMBB data experienced in the terrestrial backhaul in both networks. 

\begin{figure}[h]
    \centering
    \includegraphics[width=3.5in]{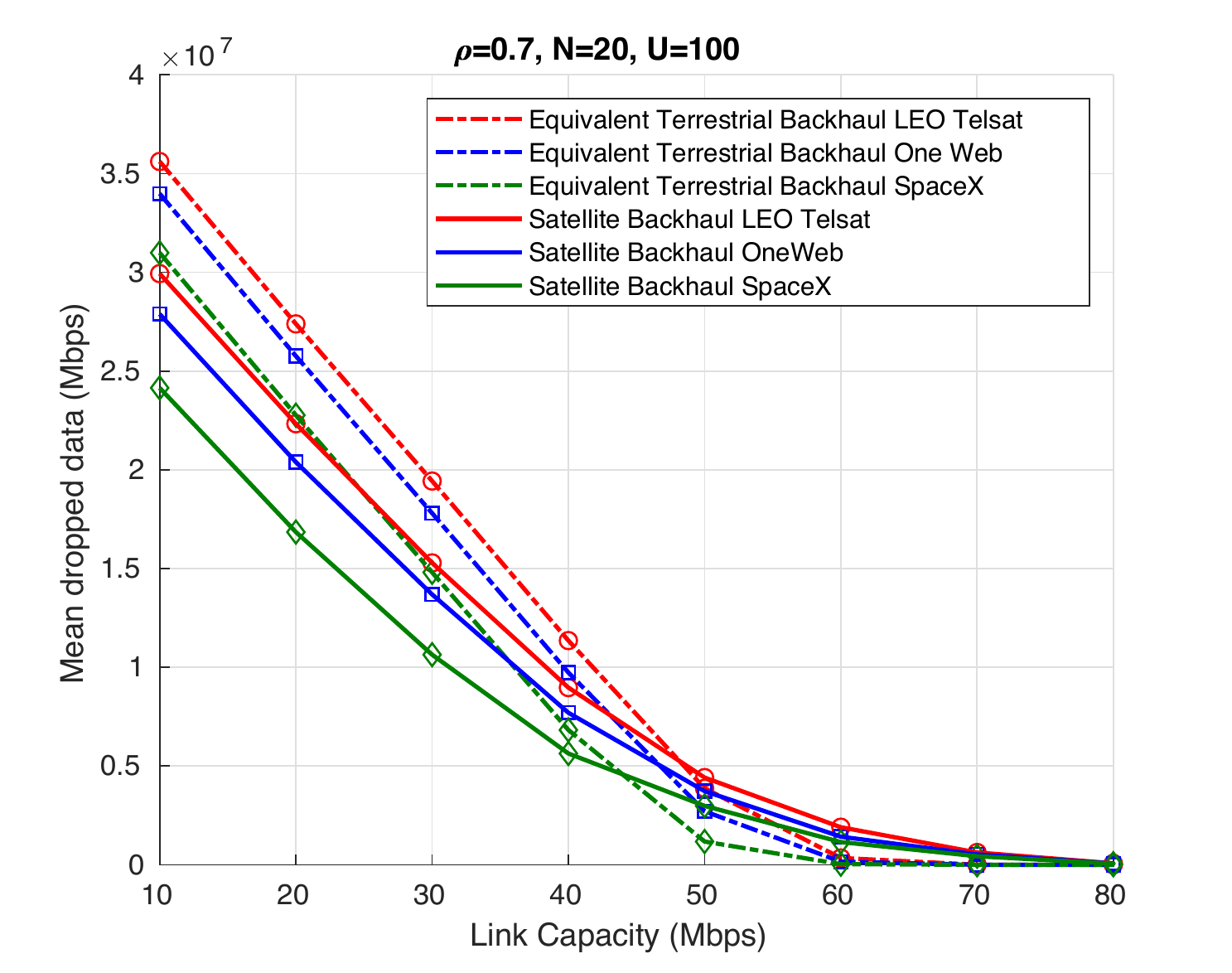}
    \caption{Mean Dropped eMBB Data as function of Link Capacity.}
    \label{fig:2}
\end{figure}

As illustrated in Fig.\ref{fig:2}, the data dropping significantly diminishes with larger link capacities in both cases. We observe also that our scheme has lower mean dropping compared to the benchmark. We note that our scheme performs better especially for the small link capacities (i.e. less than 30 Mbps).  

Afterwards, we study the network availability by assessing the percentage of the successfully transmitted data to the backhaul of the totally sent data. 

\begin{figure}[h]
    \centering
    \includegraphics[width=3.5in]{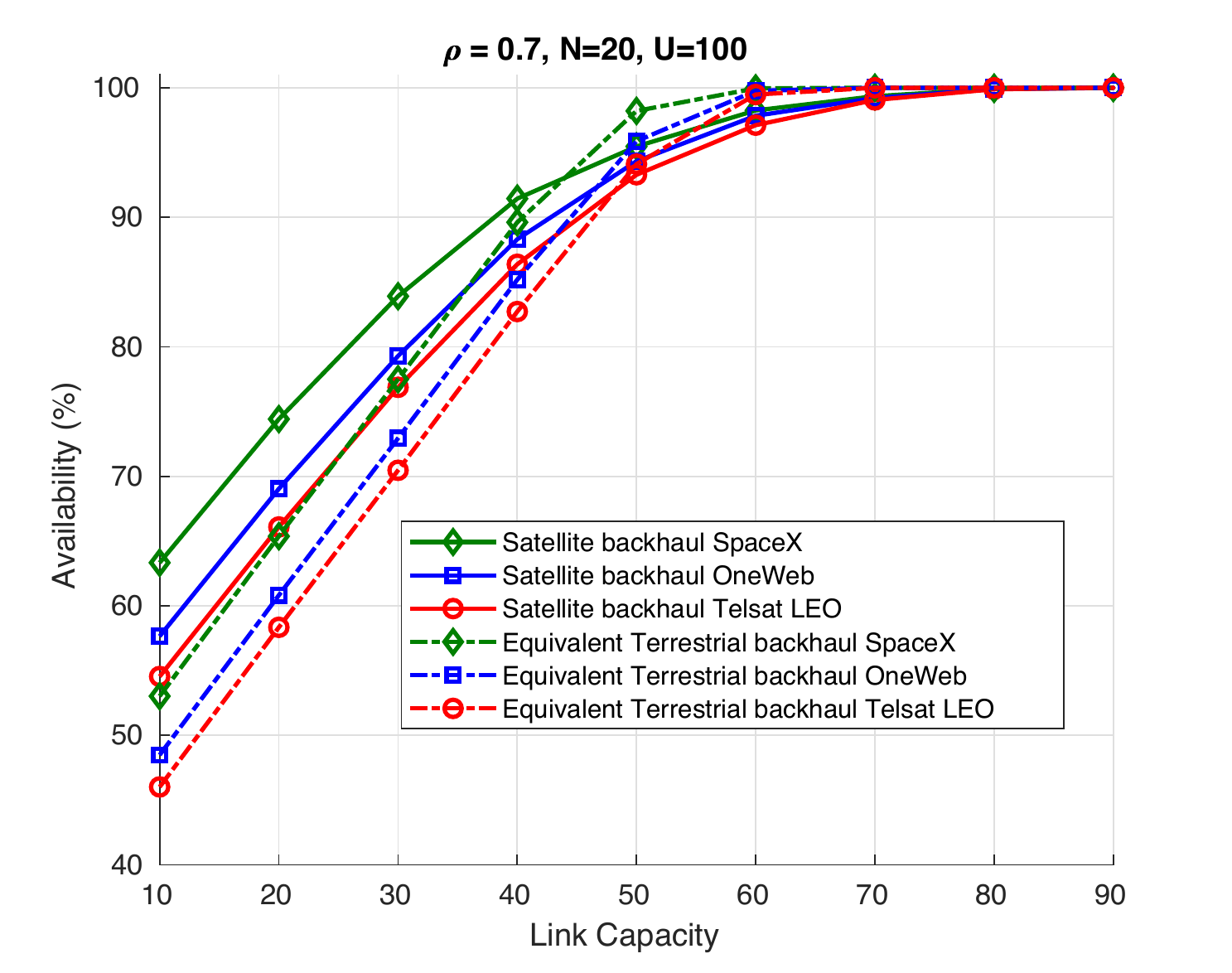}
    \caption{Mean Availability  as function of Link Capacity.}
    \label{fig:3}
\end{figure}

The results presented in Fig.\ref{fig:3} endorse the previous dropping findings. Precisely, the network availability improves with larger link capacities for both networks. We remark that our scheme  enhances the network availability particularly for the small link capacities (i.e. less than 30 Mbps) compared to the terrestrial benchmark. We notice also that the network availability stabilizes at  100\%  for the large link capacities (i.e. more than 80 Mbps). 
 
These  {simulation results} prove that ISTN outperforms the terrestrial network in terms of delay and availability when the link capacity is increased. The performance improvement of ISTN is remarkable especially in dense networks (link capacities less than 30 Mbps). However, increasing the link capacity beyond a certain limit (e.g. 80 Mbps in our case) is not always efficient.
   
   \begin{figure}[h]
    \centering
    \includegraphics[width=3.5in]{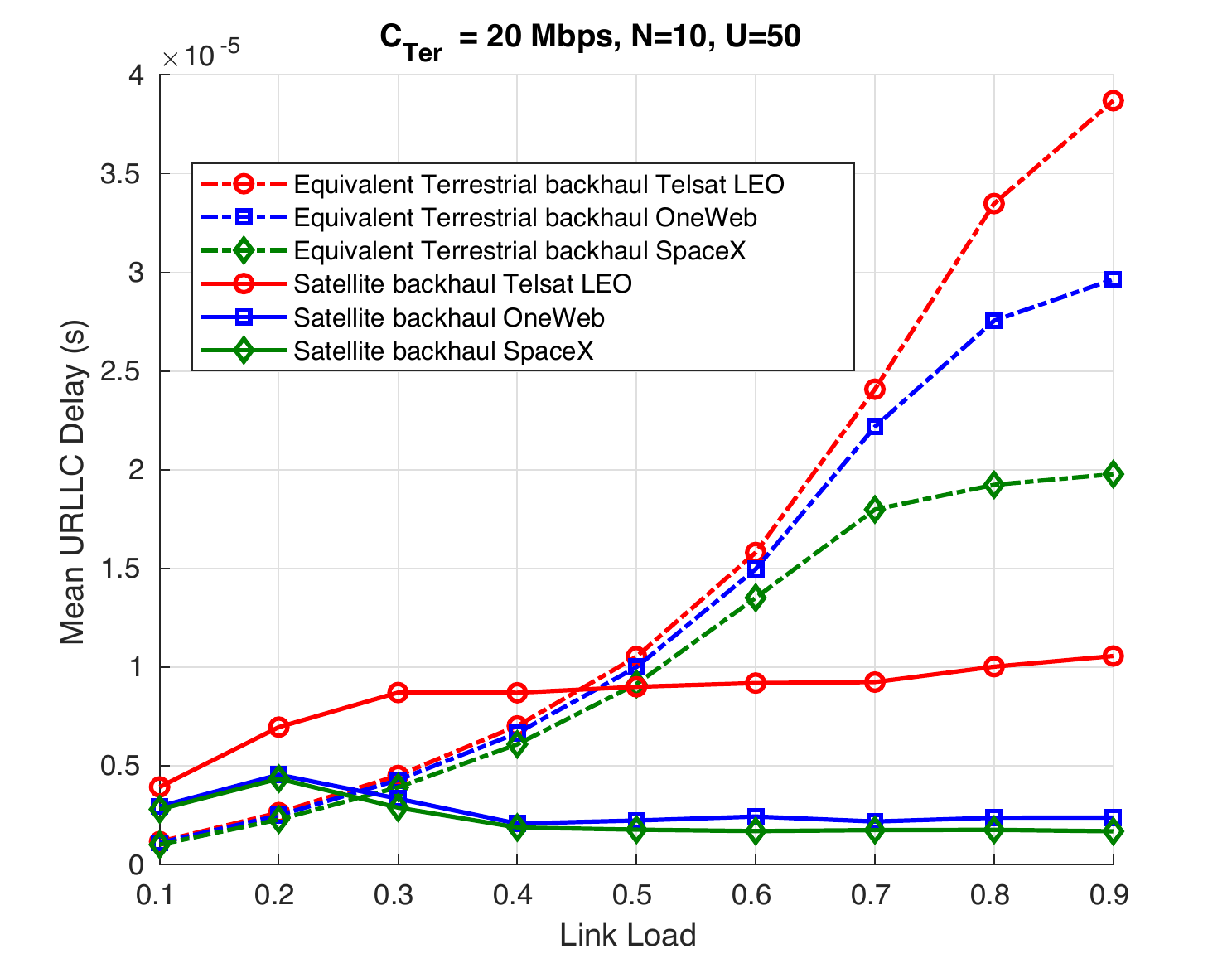}
    \caption{Mean URLLC Delay as function of Link Load.}
    \label{fig:7}
\end{figure}
In  the  second  simulation,  we inspect the influence of the network load on both networks performance. We  vary  the  link load $\rho$ between 0.1 and 0.9 for a fixed network capacity ${C}_{\mathrm{Ter}} = 20$ Mbps that characterizes dense networks. We study first the variation of the URLLC delay experienced in the terrestrial backhaul in both networks.

The results depicted in Fig. \ref{fig:7} show that our scheme surpasses the benchmark starting from a link load equal to 0.5 for Telsat LEO system and 0.3 for OneWeb and SpaceX respectively. The improvement is especially important for $\rho=0.9$. This finding underlines that our scheme helps to reduce dramatically the delays  in dense networks.

Then, we study the mean amount of the dropped eMBB data experienced in the terrestrial backhaul in both networks as depicted in Fig. \ref{fig:6} .
  
 \begin{figure}[h]
    \centering
    \includegraphics[width=3.5in]{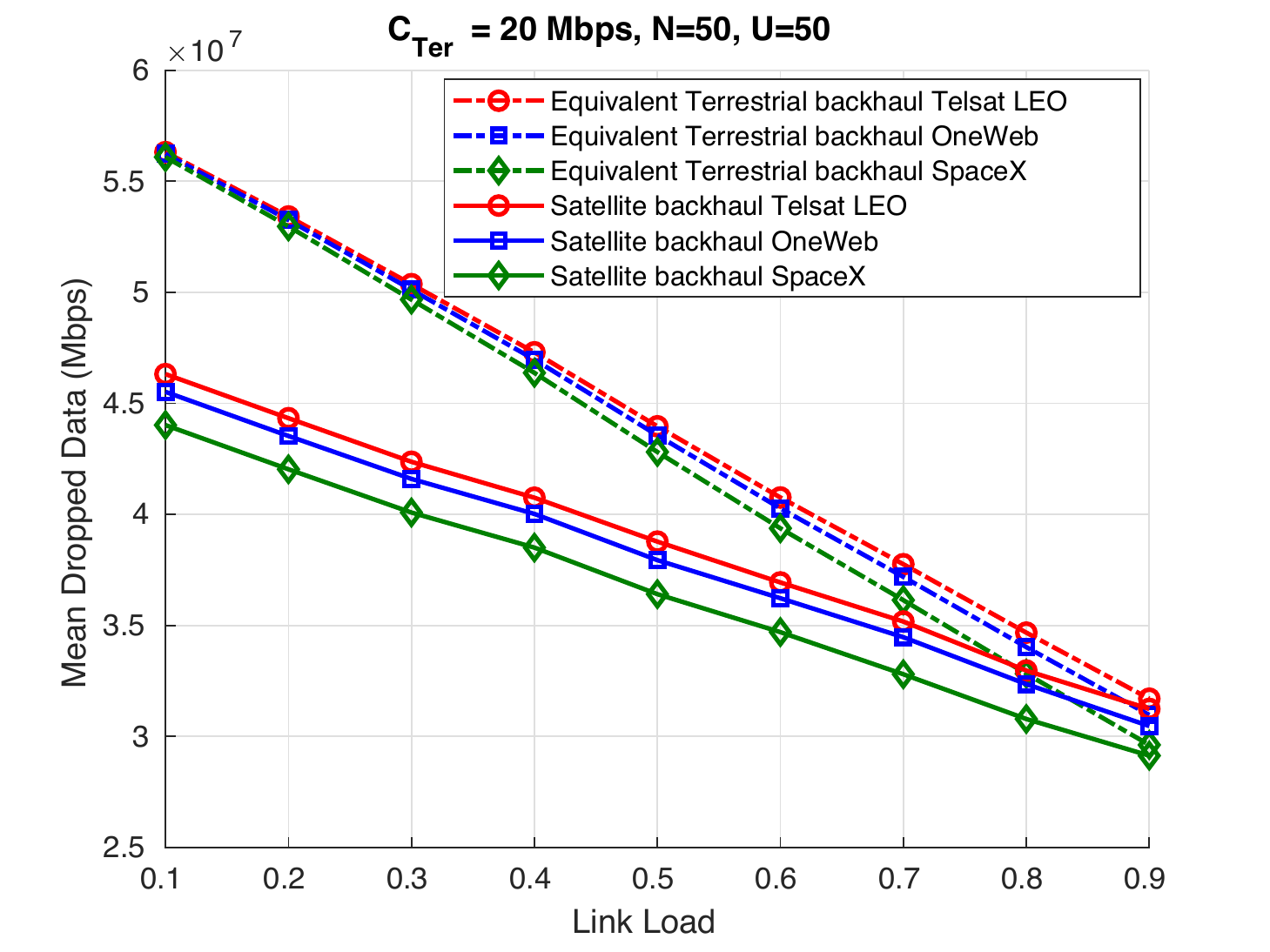}
    \caption{Mean Dropped eMBB Data as function of Link Load.}
    \label{fig:6}
\end{figure}

We notice that the mean dropped eMBB traffic decreases when the network load increases. This observation is due to the fact that a more important traffic volume is accepted in the network when the load is higher. 
We note also that the mean dropped eMBB traffic is higher for the terrestrial benchmark compared to our scheme mainly for low network loads (i.e. less than 0.3).  This observation is due to the fact that more eMBB packets are rejected to maintain a low load in the network.

 \begin{figure}[h]
    \centering
    \includegraphics[width=3.5in]{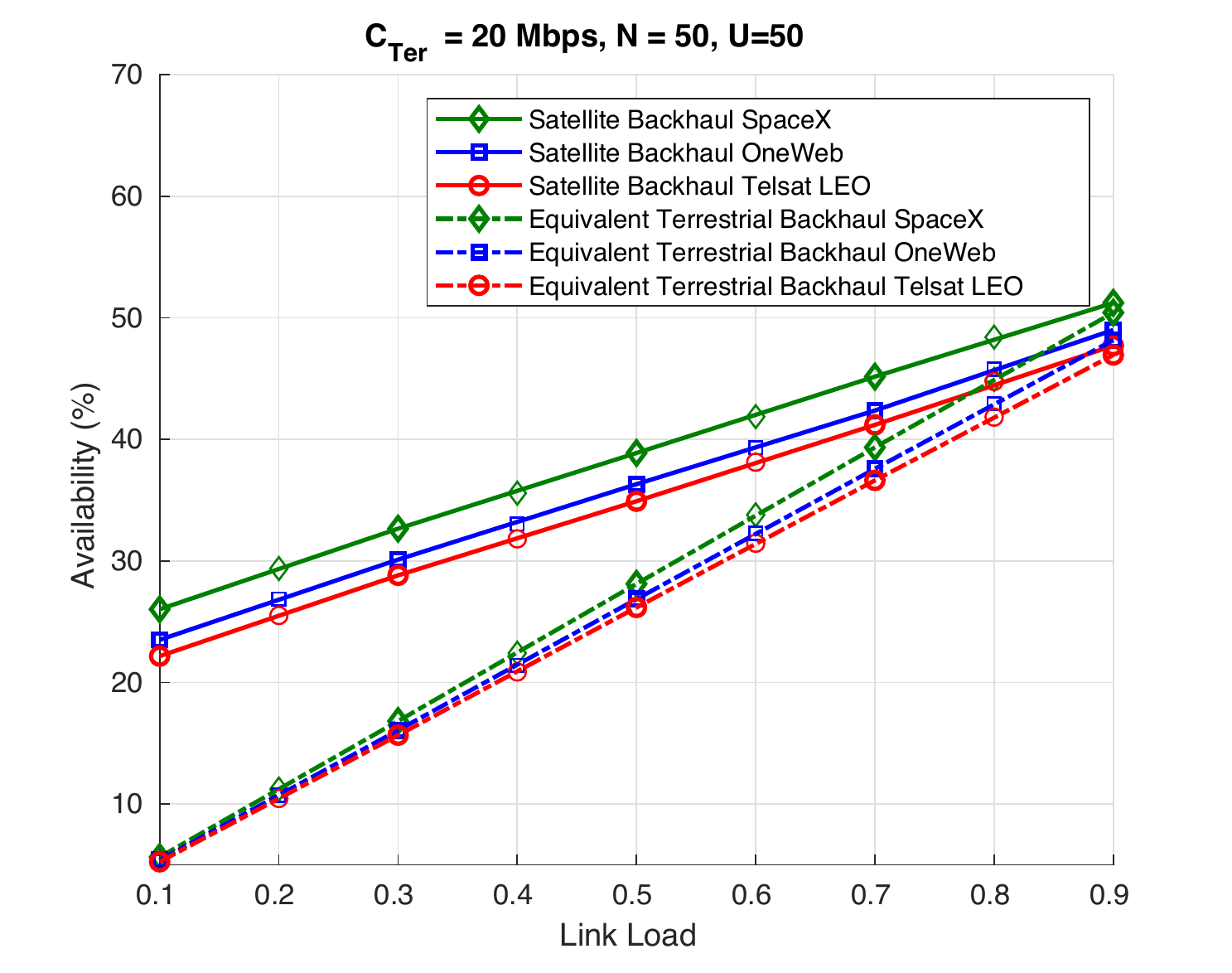}
    \caption{Mean Availability as function of Link Load.}
    \label{fig:5}
\end{figure}

Afterwards, we study the network availability.  The results presented in Fig.\ref{fig:5} show that the network availability improves with larger network loads in both cases. We remark that our scheme  enhances the network availability particularly for the low loads (i.e. less 0.3) compared to the terrestrial benchmark. This observation is due to the fact that more packets are dropped when the loads are lower. 
 
Both simulation results indicate  that a better performance is reached when the  constellation  features are more robust. Indeed, the best results are obtained  when our scheme uses SpaceX parameters, which include a higher capacity and a stronger carrier to noise density  (c.f. Table II).  

In the third simulation, we evaluate experimentally the  CDF  of the waiting time to endorse the previous results obtained through the mean values.  We compare this experimental CDF to the analytical function derived in equation (\ref{eq:36}) for both networks by using the  features of the constellation Telsat.
\begin{figure}[h]
    \centering
    \includegraphics[width=3.5in]{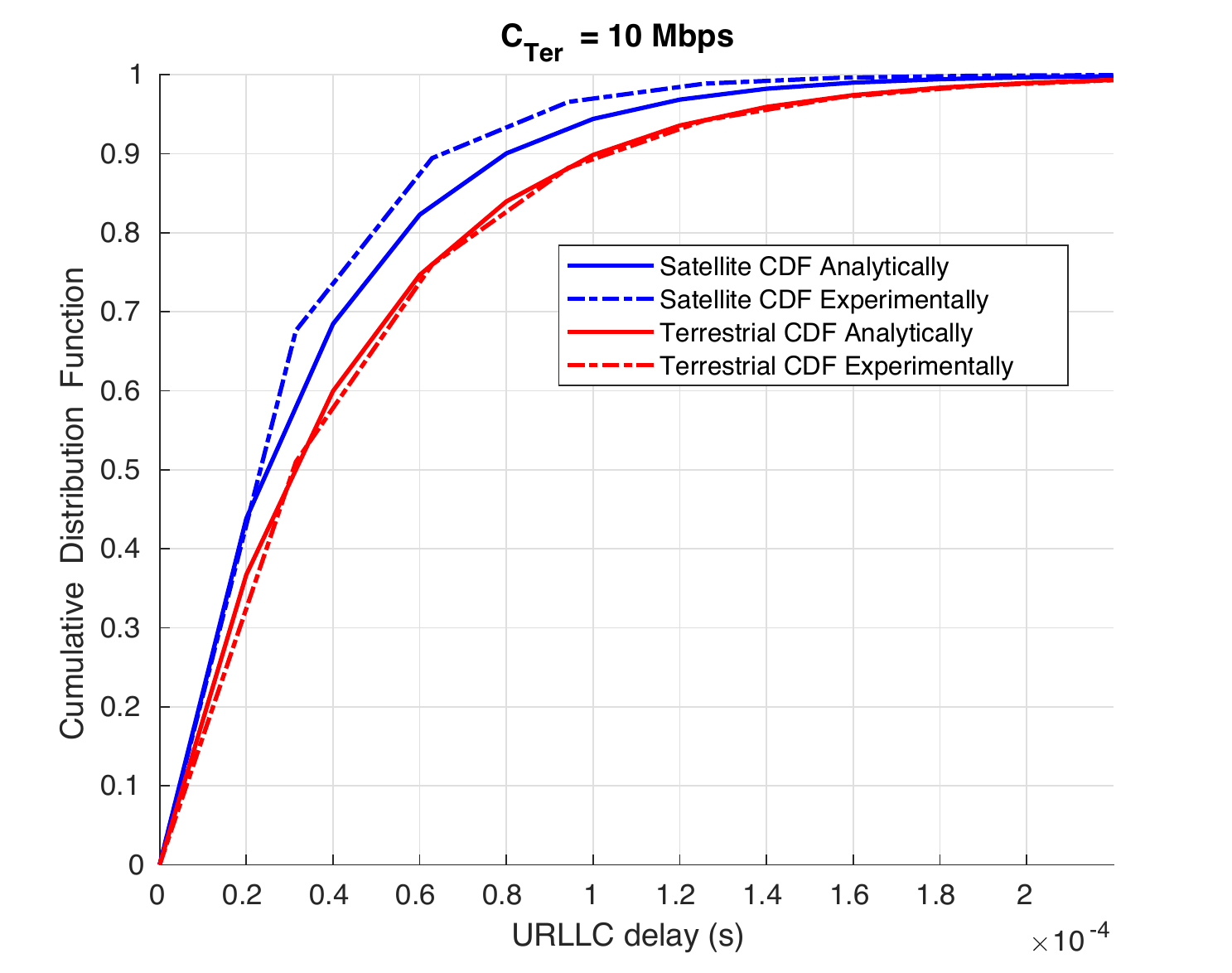}
    \caption{URLLC Delay Cumulative Distribution Function}
    \label{fig:8}
\end{figure}

As depicted in Fig. \ref{fig:8}, we note first the strong agreements between the analytical and the experimental distributions for ISTN and for the terrestrial network. We note also that the CDF curve of ISTN is above the CDF curve of the terrestrial network. This observation can be interpreted mathematically as
\begin{equation}
    \Pr(D_{\mathrm{ISTN}} < D) \geq \Pr(D_{\mathrm{Ter}}<D)
\end{equation}
where $D$, $D_{\mathrm{ISTN}}$ and $D_{\mathrm{Ter}}$ are respectively a fixed delay value, the delay experienced in ISTN and the delay experienced in the terrestrial network. This finding  {emphasizes} that the integration of the satellite network improves the delay experienced in the terrestrial bakchaul.

\section{Conclusion}
In this paper, we proposed an offloading scheme in ISTN to steer the different types of traffics to the transport network that meets their transmission requirements. In particular, we dealt with URLLC and eMBB traffics and offload the former to the terrestrial backhaul, while the latter is offloaded to the satellite network. The proposed scheme used the satellite link efficiently to accommodate high eMBB traffic while prioritizing the URLLC traffic. Our findings indicated that the integration of satellite networks with terrestrial networks is useful to relieve the capacity limitations in terrestrial backhaul. Precisely, our scheme succeeded in decreasing the dropped amount of eMBB traffic significantly in the terrestrial backhaul. Thus, the eMBB applications became able to transmit their large payload successfully to the core network.
Moreover, our scheme reduced the transmission delay experienced by the URLLC packets in the terrestrial backhaul to meet the applications' low latency requirements. Hence, it also improves the network availability for broader access to new users while satisfying their different QoS requirements. Our future work will be orientated towards machine learning in ISTN to predict the appropriate offloading times to anticipate packets dropping and enhance network availability.

\bibliographystyle{abbrv}
\bibliography{main}
\vspace{1cm}
\newpage

\begin{wrapfigure}{l}{25mm} 
    \includegraphics[width=5in,height=1.25in,clip,keepaspectratio]{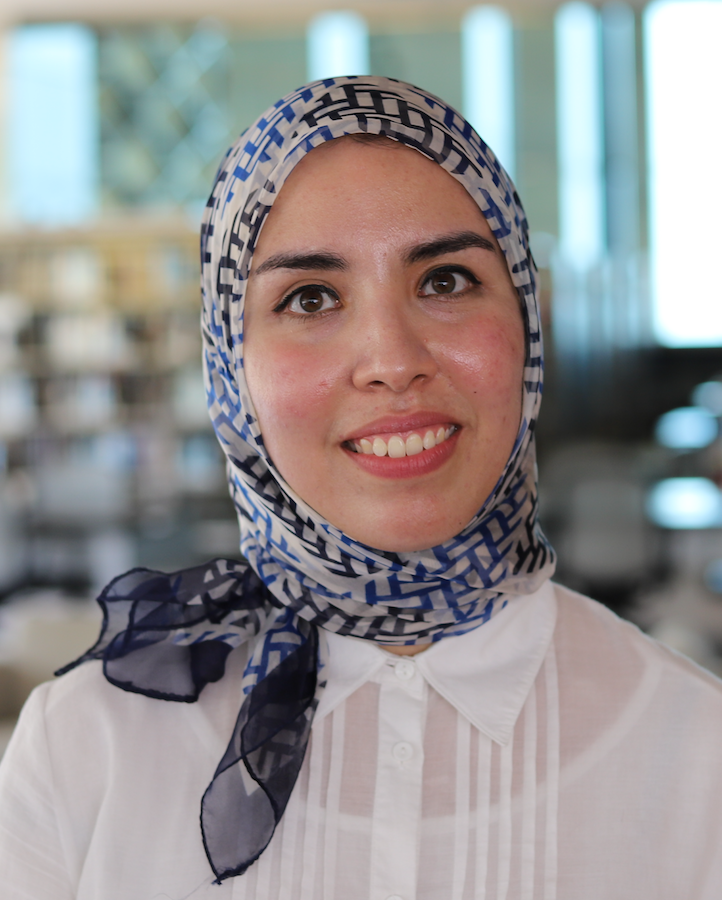}
  \end{wrapfigure}\par
  \textbf{Wiem Abderrahim} (S'14 - M'18) accomplished her undergraduate studies in electrical engineering at the Higher School of Communications of Tunis, Carthage University, Tunisia in 2013. She received her Doctoral Degree in Information and Communication Technologies from the same university in 2017. She worked as a lecturer and then as an adjunct professor at the Higher School of Communications of Tunis between 2014 and 2018. Currently, she is a postdoctoral fellow within King Abdullah University of Science and Technology (KAUST), Thuwal, Saudi Arabia. Her research interests include cloud computing,  network virtualization and recently satellite communications and machine learning.  \\
  \begin{wrapfigure}{l}{25mm} 
    \includegraphics[width=5in,height=1.25in,clip,keepaspectratio]{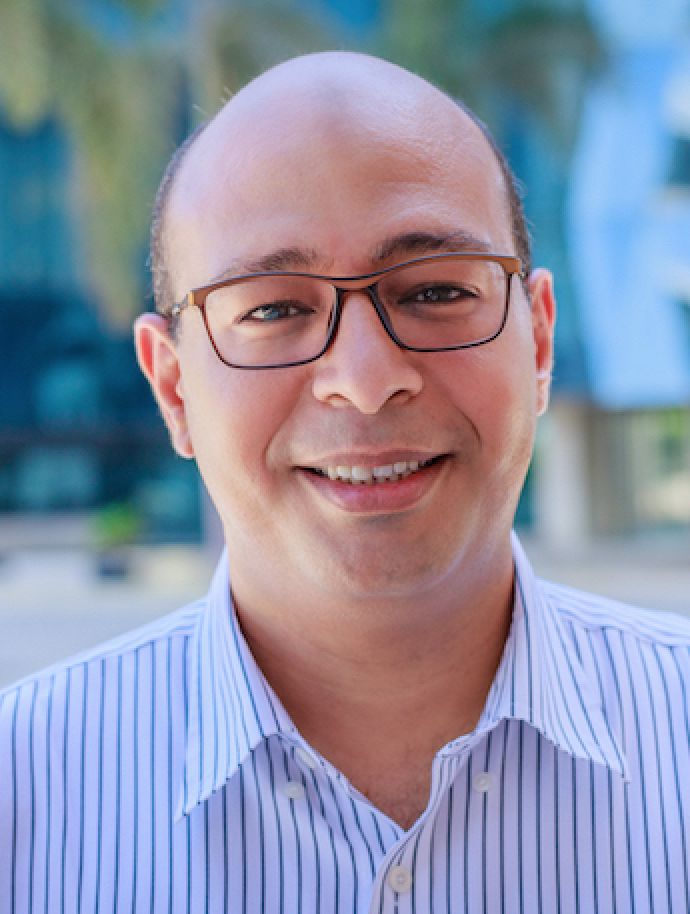}
  \end{wrapfigure}\par
  \textbf{Osama Amin} (S'07, M'11, SM'15) received his B.Sc. degree in electrical and electronic engineering from Aswan University, Egypt, in 2000, his M.Sc. degree in electrical and electronic engineering from Assiut University, Egypt, in 2004, and his Ph.D. degree in electrical and computer engineering, University of Waterloo, Canada, in 2010. In June 2012, he joined Assiut University as an assistant professor in the Electrical and Electronics Engineering Department. Currently, he is a research scientist in the CEMSE Division at KAUST, Thuwal, Makkah Province, Saudi Arabia. His general research interests lie in communication systems and signal processing for communications with special emphasis on wireless applications.  \\
 
 \begin{wrapfigure}{l}{25mm} 
    \includegraphics[width=5in,height=1.25in,clip,keepaspectratio]{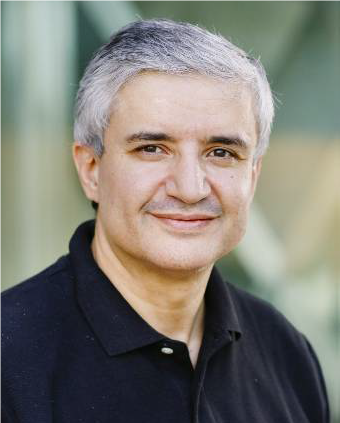}
  \end{wrapfigure}\par  \textbf{Mohamed Slim Alouini} (F'09) was born in Tunis, Tunisia. He received the Ph.D. degree in Electrical Engineering
from the California Institute of Technology (Caltech), Pasadena, 
CA, USA, in 1998. He served as a faculty member in the University of Minnesota,
Minneapolis, MN, USA, then in the Texas A{\&}M University at Qatar,
Education City, Doha, Qatar before joining King Abdullah University of
Science and Technology (KAUST), Thuwal, Makkah Province, Saudi
Arabia as a Professor of Electrical Engineering in 2009. His current
research interests include the modeling, design, and
performance analysis of wireless communication systems.  \\

\begin{wrapfigure}{l}{28mm} 
    \includegraphics[width=5in,height=1.25in,clip,keepaspectratio]{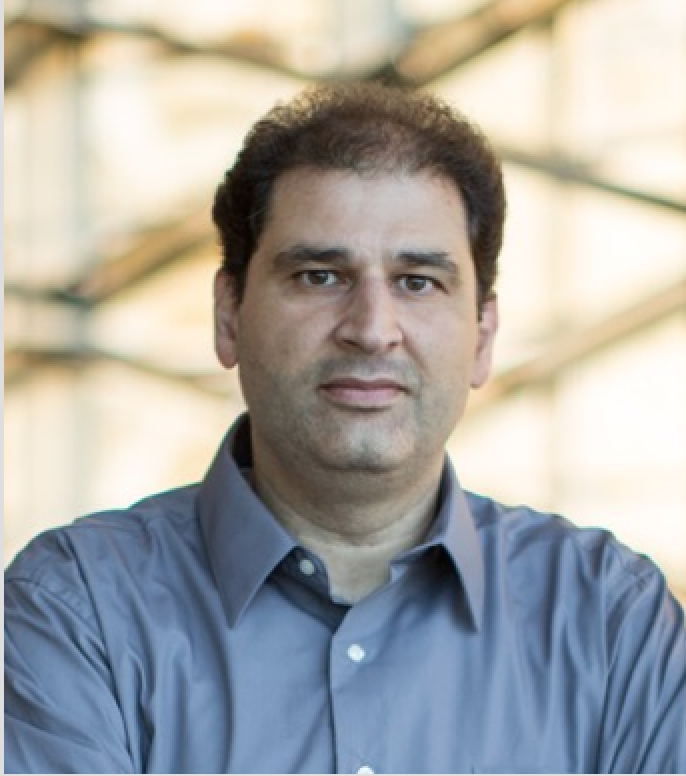}
  \end{wrapfigure}\par
  \textbf{Basem Shihada} (SM'12)  is an associate and founding professor in the Computer, Electrical and Mathematical Sciences and Engineering (CEMSE) Division at King Abdullah University of Science and Technology (KAUST). He obtained his PhD in Computer Science from University of Waterloo. In 2009, he was appointed as visiting faculty in the Department of Computer Science, Stanford University. In 2012, he was elevated to the rank of Senior Member of IEEE. His current research covers a range of topics in energy and resource allocation in wired and wireless networks, software defined networking, internet of things, data networks, smart systems, network security, and cloud/fog computing.

\end{document}